# Charge redistribution and the Magnetoelastic transition across the first-order magnetic transition in (Mn,Fe)$_2$(P,Si,B)

M. Maschek[1], X. You[1], M. F. J. Boeije[1], D. Chernyshov[2], N. H. van Dijk[1] and E. Brück[1]

[1] *Fundamental Aspects of Materials and Energy, Faculty of Applied Sciences, Delft University of Technology, Mekelweg 15, 2629 JB Delft, The Netherlands*
[2] *BM01, Swiss-Norwegian Beam Lines, European Synchrotron Radiation Facility, 71 avenue des Martyrs, 38000, Grenoble, France*

We used temperature dependent high-resolution x-ray powder diffraction and magnetization measurements to investigate structural, magnetic and electronic degrees of freedom across the ferromagnetic magneto-elastic phase transition in Mn$_1$Fe$_1$P$_{0.6-w}$Si$_{0.4}$B$_w$ ($w$ = 0, 0.02, 0.04, 0.06, 0.08). The magnetic transition was gradually tuned from a strong first-order ($w$ = 0) towards a second-order magnetic transition by substituting P by B. Increasing the B content leads to a systematic increase in the magnetic transition temperature and a decrease in thermal hysteresis, which completely vanishes for $w$ = 0.08. Furthermore, the largest changes in lattice parameter across the magnetic transition occur for $w$ = 0, which systematically becomes smaller approaching the samples with $w$ = 0.08. Electron density plots show a strong directional preference of the electronic distribution on the Fe site, which indicates the forming of bonds between Fe atoms and Fe and P/Si in the paramagnetic phase. On the other hand, the Mn-site shows no preferred directions resembling the behaviour of a free electron gas. Due to the low B concentrations ($w$ = 0 - 0.08), distortions of the lattice are limited. However, even small amounts of B strongly disturb the overall topology of the electron density across the unit cell. Samples containing B show a strongly reduced variation in the electron density compared to the parent compound without B.

## I. INTRODUCTION

Growing energy demands and declining natural resources require the development of sustainable, clean and energy-efficient technologies. Magnetocaloric materials have great potential to efficiently exchange thermal energy and magnetic energy, because they generate heat when placed in a magnetic field. Upon removal of the field, the material cools down below the environmental temperature. Magnetic cooling and magnetic heat pumps are promising techniques due to their increased efficiencies, which can be 20 to 30% higher compared to conventional technologies [1-7]. In the reversed process, a temperature-induced magnetization change can be utilized to convert low-temperature waste heat (< 80 °C) into mechanical energy by thermomagnetic motors [8-14], which may further be converted into electricity. The utilization of the magnetocaloric effect (MCE) near room temperature could reduce the worldwide energy consumption and lower the emission of greenhouse gasses, due to significantly higher efficiencies and the use of environmentally friendly and abundantly available base materials.
The magnitude of the MCE crucially depends on the nature of the phase transition and the change in magnetization with temperature. A first-order magnetic transition (FOMT) is typically associated with a large jump in the magnetization and a discontinuous change in lattice parameters and volume across the Curie-temperature $T_C$. Furthermore, a FOMT is generally accompanied by latent heat and thermal hysteresis. The MCE is maximized and results in an enhanced adiabatic temperature change in the form of a large latent heat, which is beneficial for magnetic cooling devices. However, the volume change across $T_C$ in a FOMT may damage the structural integrity of the material after multiple cycles and reduce longevity. On the other hand, a second-order magnetic transition (SOMT) does not involve a discontinuous jump in volume and is not accompanied by latent heat or hysteresis. Typically the change in magnetization is extended over a wider temperature range in a SOMT compared to a FOMT.

Close to a magnetic phase transition the (reversible) change in Gibbs free energy per unit volume $\Delta G$ can be described by the Landau-model [15]:

$$\Delta G = \frac{\alpha}{2}M^2 + \frac{\beta}{4}M^4 + \frac{\gamma}{6}M^6 - \mu_0 HM \quad (1)$$

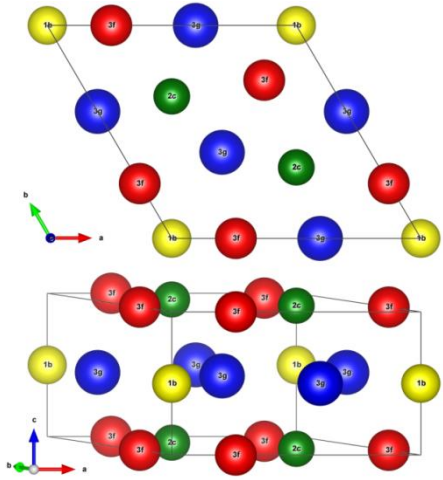

Fig. 1: Hexagonal unit cell (P-62m) of (Mn, Fe)$_2$(P, Si, B). Fe and Mn preferably occupy the *3f* and *3g* Wykoff positions, respectively. The *1b* and *2c* positions are occupied by P and Si. B substitutes P on the *1b* position.

where $M$ is the magnetization and $\mu_0 H$ the applied magnetic field. Parameter $\alpha = \alpha_0(T - T_0)$ depends linearly on temperature with respect to a characteristic temperature $T_0$. Parameters $\alpha_0 > 0$, $\beta$, $\gamma > 0$ are temperature-independent constants. Minimization with respect to the magnetization $M$ ($\partial \Delta G / \partial M = 0$) leads to the equation of state:

$$\alpha + \beta M^2 + \gamma M^4 = \frac{\mu_0 H}{M} \qquad (2)$$

The parameter $\beta$ can be negative or positive describing a FOMT or SOMT, respectively. Fine-tuning the character of the magnetic transition towards the critical point (CPT) between FOMT and SOMT ($\beta = 0$) could be an essentials step to achieve the best compromise of maximizing the efficiency for applications while keeping a reversible process to ensure longevity of the MCE material.

Promising MCE materials are (Mn,Fe)$_2$(P,Si,B) compounds [3], which combine tuneable Curie temperatures and large magnetization changes. (Mn,Fe)$_2$(P,Si,B) crystallizes in a hexagonal crystal structure (space group $P\bar{6}2m$) with two layers, where the magnetic atoms Fe and Mn preferably occupy the crystallographic *3f* and *3g* Wykoff position, respectively (Fig. 1) (drawn by VESTA [16]). The non-magnetic atoms P and Si occupy the *1b* and *2c* site. B preferably replaces P on the *1b* site [17].

The MCE in (Mn,Fe)$_2$(P,Si,B) compounds originates from the interplay between the two magnetic layers with magnetic atoms on the crystallographic *3g* and *3f* site, respectively. This so-called "mixed magnetism" [3] is governed by *strong magnetism* (*3g* site) and *weak magnetism* (*3f* site). The magnetic moment on the *3g* site stays relatively stable across the transition, whereas the moment on the *3f* site is strongly reduced in magnitude when the paramagnetic phase is entered. This strong reduction in the magnetic moment on the *3f* site is attributed to covalent bond formation with the non-magnetic atoms in the paramagnetic state [18,19]. The mechanism driving the character of the ferromagnetic magneto-elastic transition, and consequently the magnetic entropy change, is still not fully understood in these compounds as systematic studies on the charge redistribution across the ferromagnetic transition with samples gradually changing from a FOMT towards a SOMT were unavailable.

Our main goal is to explore the driving forces in (Mn,Fe)$_2$(P,Si,B) compounds that define the character of the magneto-elastic transition. Therefore, we gradually tuned the magnetic transition from a strong first-order towards a second-order character by incrementally substituting P by B. We studied the correlations between magnetic and structural properties across $T_C$ and investigated how the electron density redistribution develops across $T_C$, when the magnetic character is tuned stepwise from a FOMT towards a SOMT. Our investigation aims at identifying the nature of the magnetic phase transition to be able to better design suitable MCE materials for magnetic cooling, heat pumps and magnetic energy conversion. Our interest is to identify the essential parameters controlling the strength of the MCE and consequently the change in magnetization across the transition.

## II. EXPERIMENT

The five studied samples were Mn$_1$Fe$_1$P$_{0.6-w}$Si$_{0.4}$B$_w$ with $w$ = 0, 0.02, 0.04, 0.06 and 0.08. We mixed powders of Mn, Fe, P, Si and B via ball milling for 10 hours at 380 rounds per minute. The resulting powders were pressed into cylindrical tablets, which were sealed in quartz tubes in Ar atmosphere. Temperature-dependent magnetization was measured in a magnetometer equipped with a superconducting quantum interference device (SQUID, Quantum Design MPMS 5XL). We determined magnetic entropy changes by field-dependent isothermal magnetization measurements utilizing the Maxwell relations. High-resolution x-ray diffraction was performed at the BM01A beamline at the European Synchrotron Radiation Facility (ESRF) using a wavelength of 0.69264 Å. The measurement covered a temperature range from 150 to 500 K in temperature steps of 2 K. The temperature was controlled by a Nitrogen Cryostream. The samples were put in capillaries with 0.5 mm diameter and spun. Diffraction data were analysed by Rietveld refinement using the FullProf Suite [20]. Electron densities were calculated by VESTA [16] and processed in Matlab [21].

## III. RESULTS

### A. Magnetic properties

The temperature dependent magnetization of $Mn_1Fe_1P_{0.6-w}Si_{0.4}B_w$ was measured in a magnetic field of 1 T and shows increasing $T_C$'s (heating) ranging from 281 to 330 K for $w = 0$ and $w = 0.08$, respectively (Fig. 2(a)). The largest thermal hysteresis of $\Delta T_{hys} = 54$ K is observed for the sample with no B ($w = 0$). The hysteresis gradually decreases with increasing B content and practically disappears in the sample with $w = 0.08$.

The magnetic entropy change $-\Delta S$ was determined in magnetic fields up to 2 T (Fig. 2(b)). In a magnetic field of 1 T $Mn_1Fe_1P_{0.6}Si_{0.4}$ shows a maximum of $-\Delta S = 6.7$ $Jkg^{-1}K^{-1}$ at $T_C$. An increasing B content leads to an increase in magnetic entropy change with a maximum of $-\Delta S = 13.1$ $Jkg^{-1}K^{-1}$ for $w = 0.06$. For $w = 0.08$ the entropy change slightly drops to $-\Delta S = 6.8$ $Jkg^{-1}K^{-1}$.

### B. Structural properties

The intensity distribution for scattering angles $2\theta$ from 17° to 21° as a function of temperature are shown in Figs. 3(a)-3(e). The peak intensities of the main peaks are represented in a color-coded contour plot. We observe clear discontinuous jumps in the diffraction peaks across the ferromagnetic transition for the samples with $w = 0 - 0.06$ (e.g. see the {210} diffraction peak). Furthermore, we observe a small temperature interval, where ferromagnetic and paramagnetic phase coexist. Both observations are characteristic for a material with a FOMT character. The weak peak close to 20° corresponds to the {220} reflection of the impurity phase $(Mn,Fe)_3Si$. This phase occurs in all samples, however only in small amounts (< 10 %). With increasing B content the discontinuous steps in the scattering angle of the main reflections at $T_C$ become smaller and are nearly continuous for the sample with largest B content ($w = 0.08$). Such a behaviour indicates a weak FOMT (close to the critical point) being on the verge of a SOMT. Figs. 3(f)-3(j) shows the corresponding low temperature data and Rietveld refinements for all samples at 150 K.

The obtained changes in lattice parameters and unit-cell volume with temperature are shown in Figs. 3(k)-3(n). The in-plane lattice parameter $a$ shows a jump and decreases upon heating across $T_C$ (Fig. 3(k)). On the other hand, the out-of-plane lattice parameter $c$ shows the opposite behaviour and exhibits a steep increase upon heating across $T_C$

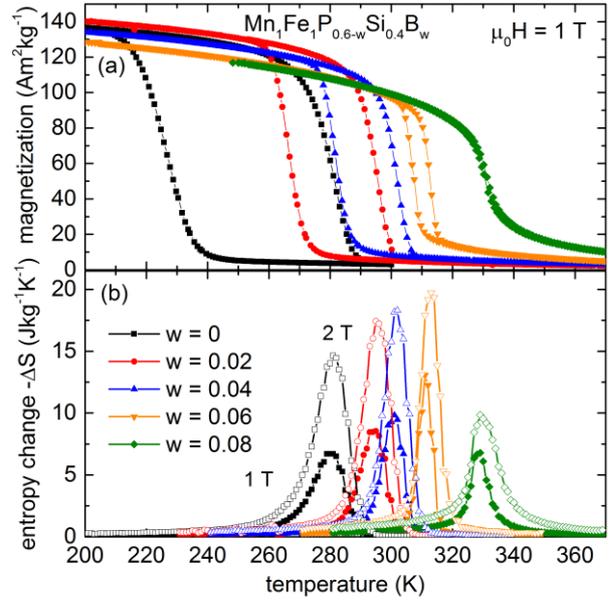

fig. 2: (a) Temperature dependent magnetization for $Mn_1Fe_1P_{0.6-w}Si_{0.4}B_w$ ($w = 0, 0.02, 0.04, 0.06$ and $0.08$) obtained in a magnetic field of $\mu_0 H = 1$ T and (b) magnetic entropy changes $-\Delta S$ for 1 T (filled symbols) and 2 T (open symbols)

(Fig. 3(l)). The $c/a$ ratio as a function of temperature increases with increasing B content (Fig. 3(m)). The jumps across $T_C$ in lattice parameters are largest in the sample with $w = 0$ and gradually become smaller with increasing B content.

In addition to the main phase $Mn_1Fe_1P_{0.6-w}Si_{0.4}B_w$ we detect the $(Mn,Fe)_3(P,Si)$ impurity phase in all samples. The amount of impurity phase gradually increases with increasing B content from 6% for $w = 0$ and reaches a maximum value of 10% for $w = 0.06$.

### C. Internal coordinates of Fe and Mn

We observe shifts in the internal coordinates of the *3g* and *3f* Wykoff positions within the unit cell across $T_C$ (Fig. 4). These sites are preferably occupied by Mn (*3g*) and Fe (*3f*). The largest discontinuous shift upon heating are observed in the sample without B ($w = 0$). With increasing B content the jumps in internal coordinates at $T_C$ both become smaller. For $w = 0.08$ we observe a gradual continuous decrease of the internal coordinates upon heating. Close to the magnetic transition the uncertainties in the Rietveld refinement increase due to the coexistence of the ferro- and paramagnetic phase leading to strong overlaps of peaks and a reduced phase fraction. The bump close to 340 K on the *3f* site for the sample with $w = 0.08$ is considered as an artefact due to uncertainties in the sequential refinement.

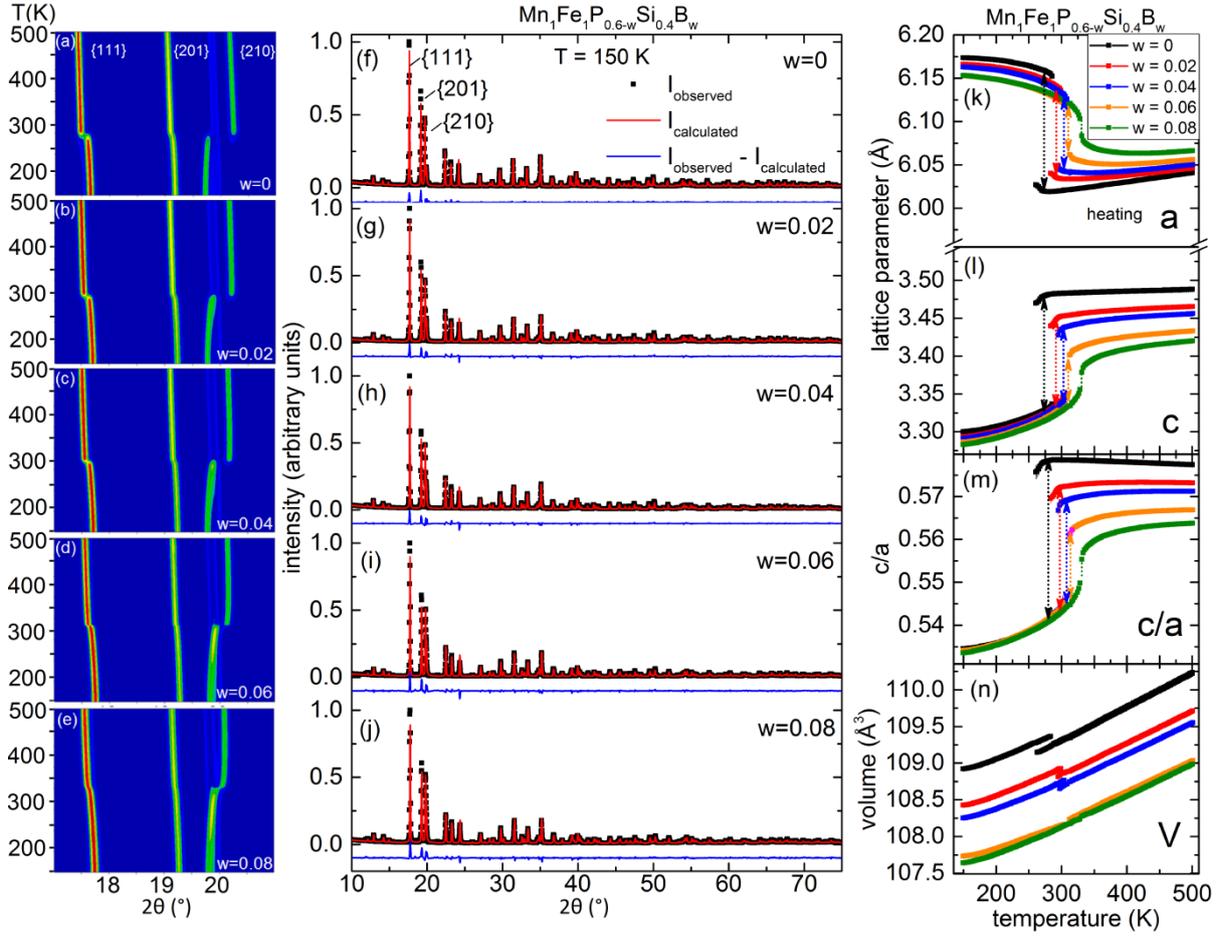

Fig. 3: (a-e) Intensity versus scattering angle $2\Theta$ (°) as a function of temperature $T$ (K) for $Mn_1Fe_1P_{0.6-w}Si_{0.4}B_w$ ($w$ = 0, 0.02, 0.04, 0.06 and 0.08) obtained in high-resolution xray powder diffraction measurements. The {111}, {201} and {210} diffraction peaks are shown across the ferromagnetic-paramagnetic transition upon heating. (f-j) x-ray diffraction patterns and Rietveld refinements for $w$ = 0 – 0.08 measured at 150 K. Temperature dependence upon heating of (k-l) hexagonal lattice parameters, (m) $c/a$ ratio and (n) the unit cell volume.

### D. Electronic structure

The high-resolution diffraction data can be used to extract the electron densities from the structure factors obtained from the refinement. In order to highlight changes in the electron density across $T_C$ we created electron density difference plots by subtracting the paramagnetic from the ferromagnetic state after normalization. The unit cells were presented in relative lattice units to subtract the phases. We assumed a constant unit cell charge at all temperatures. The diffracted intensity does however decrease with increasing temperature due to increased thermal motions of the atoms. In order to maintain a constant sum of all structure factors we multiplied the structure factors by a temperature-dependent correction factor and normalized the data. The same method has been successfully applied in our previous investigations (for more details see [18,19]). Here we investigate the electron density of the *3f* and the $3g$ site for a series of samples with varying entropy change at $T_C$.

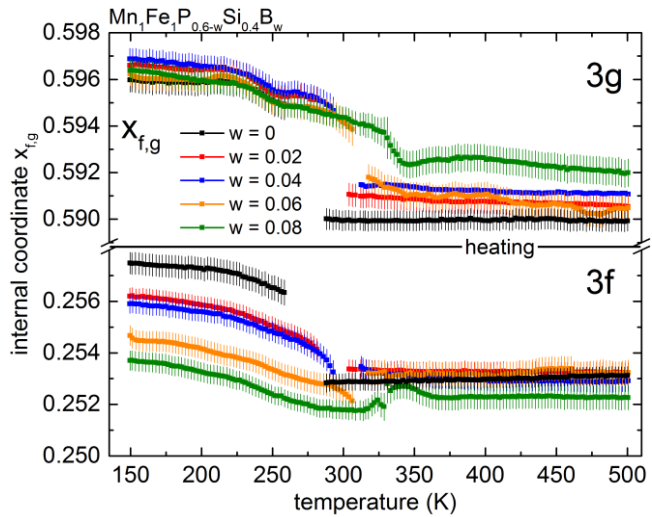

Fig. 4: Temperature dependence of the internal coordinates of the *3f* and *3g* position for $Mn_1Fe_1P_{0.6-w}Si_{0.4}B_w$ ($w$ = 0 – 0.08) across $T_C$ upon heating.

The color-coded electron density difference plots within the $a$-$b$ plane between paramagnetic (500 K) and ferromagnetic (150 K) state for $Mn_1Fe_1P_{0.6}Si_{0.4}B_w$ ($w = 0 - 0.08$) are shown for the $3f$ site (Figs. 5(a)-5(e)) and the $3g$ site (Figs. 5(f)-5(j)) containing the Fe and Mn atoms, respectively. In order to emphasize the effects originating from the different magnetic atoms, we only consider the local environment close to the $3f$ and $3g$ sites and chose different integration boundaries for the layers containing the *3f* and *3g* sites: the integration along the *c*-axis in internal coordinates of the unit cell has been performed from $z$ = -0.25 to 0.25 for the Fe-layer (*3f* site) and from $z$ = 0.25 to 0.75 for the Mn-layer (*3g* site). Furthermore, the shift in internal coordinates with temperature is removed by applying corresponding shifts given in Fig. 4, in order to emphasize on the effects originating from redistribution of electron densities across the magnetic transition.

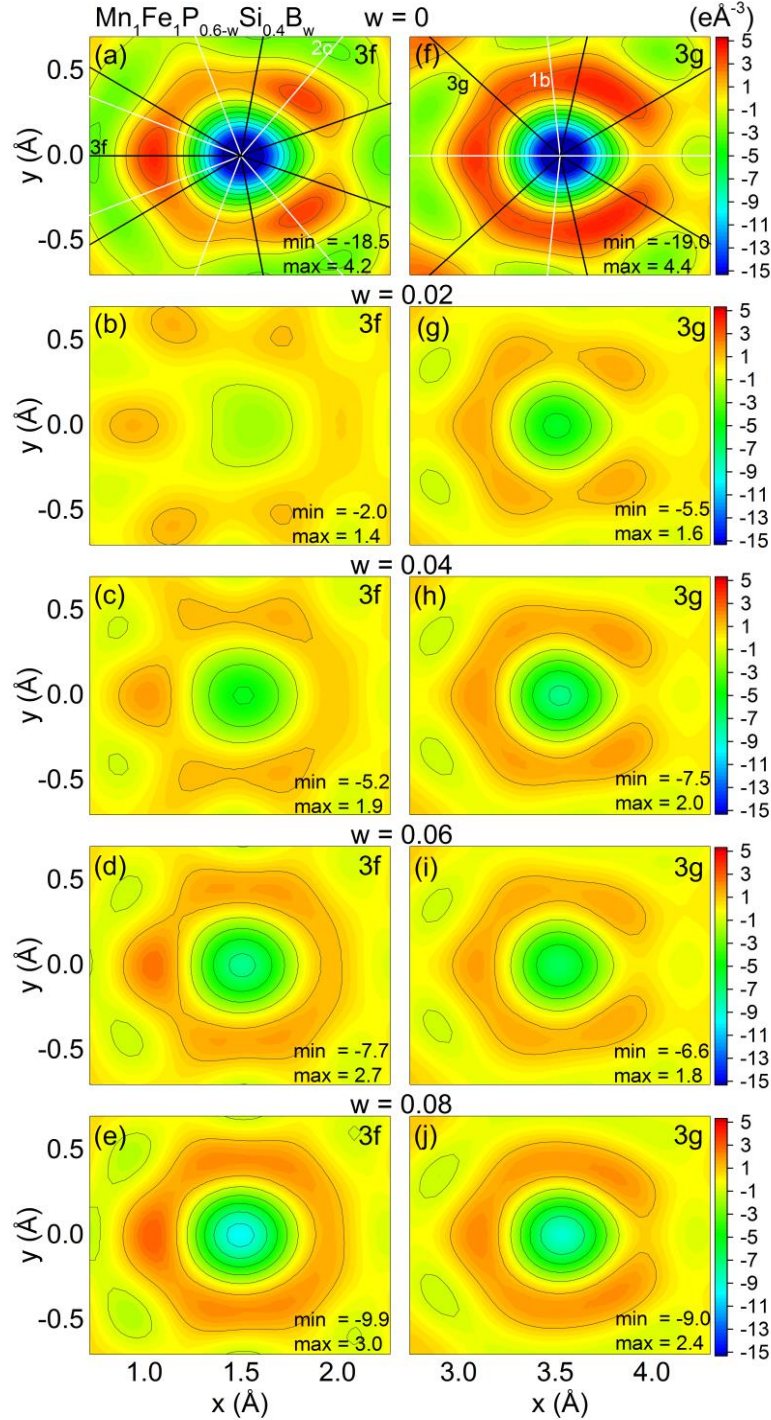

Fig. 5: Electron density difference plots between paramagnetic (500 K) and ferromagnetic (150 K) phases in the local environment of the (a-e) $3f$ and (f-j) $3g$ sites for $Mn_1Fe_1P_{0.6-w}Si_{0.4}B_w$ ($w = 0 - 0.08$). The electron densities were integrated from $z = -0.25$ to $0.25$ and $z = 0.25$ to $0.75$ (internal coordinates) for $3f$ and $3g$ positions, respectively. The shift of internal coordinates with temperature is removed by the corresponding shifts from Fig. 4. in order to emphasize the effects originating from redistribution of electron densities. The color-coded scale is given in $e\text{Å}^{-3}$.

The electron density difference between the paramagnetic and ferromagnetic state corresponds to positive (red) and negative (blue) values, respectively. We applied the same scale in $e\text{Å}^{-3}$ for all electron density difference plots for comparison. The evolution of electron density difference with changing temperature interval for $w$ = 0 (Fig. S1) and $w$ = 0.08 (Fig. S2) is given in the supplementary information.

We observe the strongest changes in the electron density around the *3f* as well as *3g* site for the sample without B ($w$ = 0) with changes ranging from -19 $e\text{Å}^{-3}$ to +4 $e\text{Å}^{-3}$. Substitution of B has a profound effect on the electronic structure and strongly weakens the electronic redistribution across the magnetic transition. The changes in the electron density difference for $w$ = 0.02 – 0.08 across $T_C$ is reduced by a factor of 3-5 compared to $w$ = 0. Furthermore, the lowest changes in the electron density difference occur in the sample with $w$ = 0.02 and then gradually increase with increasing B content. We also observe qualitative differences in the redistribution of electron density across $T_C$. The redistribution of electron density around the Fe-site (*3f*) shows strong localization in the paramagnetic state (red) pointing towards other nearest neighbour Fe-sites and the *2c* sites indicated by black lines and white lines in Fig 5(a), respectively. On the other hand, the redistribution in the paramagnetic state around the Mn site is more uniform and does not show a sizable directional character. This localization on the *3f* site and the more uniformly distributed electron density on the *3g* site in the paramagnetic state can be observed in all samples. However, electron density redistribution of the sample without B ($w$ = 0) is set apart from the other samples, due to its steep increase in electronic changes, which are much stronger compared to all other samples that contain B.

The color-coded electron density difference plots in Fig. 5 show a selected range around the magnetic atoms. By considering the close environment of one crystallographic site it is possible to remove the effect of the shift in internal coordinates (see Fig. 4) with temperature and only visualize the electronic redistribution.

We now consider the full *a-b* plane of the unit cell with same integration boundaries along the *c*-axis from $z$ = -0.25 to 0.25 for the Fe-layer (*3f* site) and from $z$ = 0.25 to 0.75 for the Mn-layer (*3g* site). We investigated the temperature dependence of the magnitude of the electronic changes across $T_C$ by determining the variance $\sigma^2$ of the electron density difference topology. In contrast to Fig. 5, this includes the shift of the magnetic atoms with temperature. Fig. 6 shows the temperature dependence of the variance of the electron density difference topology between paramagnetic and ferromagnetic phase. We fixed the reference temperature for the variance at 150 K in the ferromagnetic phase, which was subtracted from the variance obtained in the paramagnetic phase from 200 to 500 K. For comparison we plotted the temperature dependence relative to $T_C$, which is shown as vertical dashed line in Fig. 6.

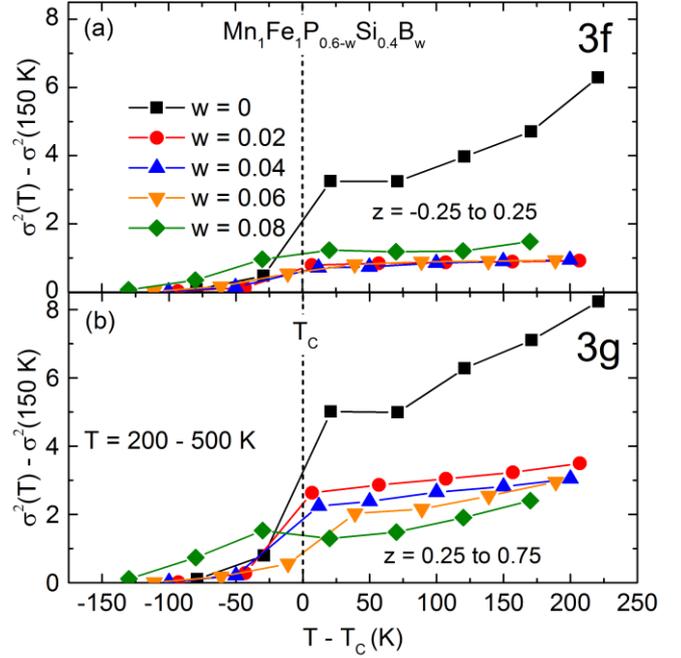

fig. 6: Variance $\sigma^2(T)$ - $\sigma^2$(150 K) of the electron density difference topology for $200 \leq T_{PM} \leq 500$ K as function of $T_{PM}$ - $T_C$ for $Mn_1Fe_1P_{0.6-w}Si_{0.4}B_w$ ($w$ = 0 – 0.08) for (a) *3f* and (b) *3g* Wykoff positions. Data includes shift of internal coordinates with temperature of the *3f* and *3g* positions. Data integration was performed in full *x-y* plane (corresponding to *a-b* plane) of the unit cell from $z$ = -0.25 to 0.25 for *3f* and $z$ = 0.25 to 0.75 for *3g* (internal coordinates). The vertical dashed line marks $T_C$.

All samples show increasing electronic changes when the magnetic transition is crossed. The strongest changes in the electron density between ferromagnetic and paramagnetic phase are observed in the sample without B ($w$ = 0). In the paramagnetic phase the electronic changes show an increasing trend up to the highest observed temperature difference. The variance observed in the sample without B ($w$ = 0) is a factor of 5 stronger on the Fe site (*3f*) and a factor of 2.5 stronger on the Mn site (*3g*) site compared to the samples with substituted B ($w = 0.02 - 0.08$). Small amounts of B already strongly supress electronic changes. The evolution observed for the samples with B ($w = 0.02 – 0.08$) show the same behaviour for the Fe site (*3f*). The changes slightly increase across $T_C$ and stay constant up to higher temperatures. For the Mn site (*3g*) the electronic changes of $w = 0.02 - 0.08$ slightly decrease with increasing B content in the paramagnetic phase and are smallest for $w = 0.08$.

## IV. DISCUSSION

In order to identify trends in our series of samples, we summarize the magnetic ($T_C$, $\Delta T_{Hys}$), structural ($\Delta(c/a)/(c/a)$) and electronic ($\sigma^2$) properties as function of B content in Fig. 7.

### A. Magnetic and structural properties

Substitution of B in $Mn_1Fe_1P_{0.6}Si_{0.4}B_w$ ($w = 0 – 0.08$) has a systematic impact on the magnetic properties. In particular, the transition gradually changes from a FOMT towards a SOMT. Decreasing discontinuous jumps in lattice parameters (Fig. 3(k)-3(l)) across $T_C$ systematically weaken the FOMT and lead to a systematic increase of the magnetic transition temperatures (Fig. 7(a)) with a decrease in hysteresis (Fig. 7(b)). The hysteresis becomes practically zero for $w = 0.08$ indicating a near SOMT character.

The gradual changes in $T_C$ and hysteresis with B substitution are also reflected in the structural properties. With increasing B content the discontinuous jump across $T_C$ in lattice parameters and $c/a$ ratio ($\Delta(c/a)/(c/a)$) gradually becomes smaller (Fig. 7(d)). Furthermore, we observe a gradual shift in internal coordinates of the Fe (*3f*) and Mn (*3g*) position between 500 and 150 K with increasing B content (Fig. 7(e)).

On the other hand, the maximum of the magnetic entropy change $-\Delta S_m$ at $T_C$ (Fig. 7(c)) does not show a similar clear trend with increasing B content. The entropy change $-\Delta S_m$ increases from $w = 0$, has a maximum at $w = 0.06$ and shows a decrease for $w = 0.08$. This behaviour resembles the trend observed in the impurity phase content. A systematic study of how the impurity phase content is related to the magnetic entropy change across the $(Mn,Fe)_2(P,Si)$ family is outside the scope of this investigation.

### B. Electron density difference plots

The electron density difference plots only reveal systematic evolution for the samples containing B ($w = 0.02 - 0.08$), where the variance of the electronic topology between 150 and 500 K does not show significant changes on the Fe (*3f*) and Mn

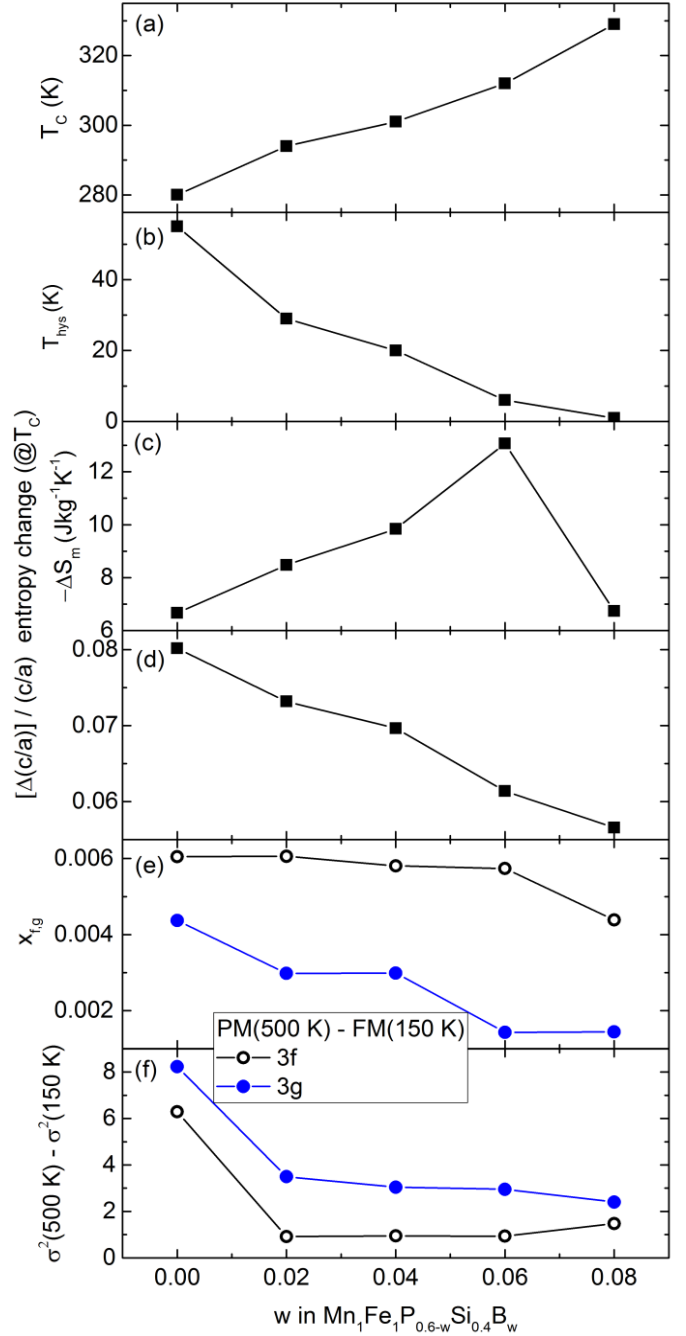

Fig. 7: (a) Curie temperature $T_C$, (b) thermal hysteresis $\Delta T_{hys}$, (c) maximum magnetic entropy change, (d) change of $c/a$ ratio in respect to 150 K $\Delta(c/a)/(c/a)$, (e) internal coordinate of the *3f* and *3g* position within the unit cell and (f) variance $\sigma^2$ of the electron density difference topology as function of Boron content $w$. The shift in internal coordinates (e) and the variance (f) of the electron density difference is shown for the difference between 500 and 150 K. Data in (f) includes the shift in internal coordinates of the *3f* and *3g* positions with temperature (shown in (e)).

(*3g*) site (Fig. 7(f)). However, we observe strongly increased variations in the electron density across $T_C$ for the sample without B ($w = 0$). Small amounts of B seem to significantly disturb the overall topology of the electron density across the whole unit cell and lead to a strongly reduced variation in the electron density compared to the parent compound without B.

The variance of the electron density difference between 150 and 500 K for $w = 0$ is a factor of six larger for *3f* and a factor of two larger for *3g* compared two all other samples. This indicates that the substitution of B and the reduction of the number of electrons in the unit cell strongly supress the electronic changes across the transition. Qualitatively all samples show a localized redistribution on the *3f* site.

However, the strong electronic changes for $w = 0$ compared to the series of samples that contain B is however not reflected in the magnetic and the structural properties.

Strong localization of the electron density in the paramagnetic state of the Fe-site pointing to other Fe atoms and the P/Si atoms indicate bond formation, which leads to a reduced magnetic moment on the Fe site above $T_C$, which has been reported in our previous investigations [19]. Furthermore, our measurements reveal that the electron density changes across $T_C$ are more evenly distributed around the Mn-site, locally resembling the behaviour of a free electron gas. The Mn site does not show indications of bond formation, which is consistent with the observation that the magnetic moment keeps its value across $T_C$. [22].

For a SOMT, the changes in the electron density, magnetic and structural properties are extended over a larger temperature range compared to a FOMT. It is expected that short-range magnetic order exists in a SOMT sample in the paramagnetic state, which continuously develops into long-range magnetic order by cooling below the magnetic transition temperature. The electron density stays relatively constant above $T_C$ in the samples with B ($w = 0.02 - 0.08$), which indicates that no significant short-range order is present. However, the sample without B ($w = 0$) exhibiting the strongest FOMT shows that the changes in the electron density difference still weakly increase with heating above $T_C$.

The changes in the electronic structure with increasing B content can be related to the mechanical properties. A previous study showed a strongly increased mechanical stability, when B was present in the sample [23]. A systematic study of the mechanical properties and the evolution of electron densities across $T_C$ with changing B content would be desirable to investigate a possible correlation between both properties.

### C. Magneto-elastic coupling

The change in lattice parameters across the magnetic transition is closely related to the elastic strain in the material. In the ferromagnetic state it is energetically favourable to induce elastic strains and thereby deform the structure. We determined the strain and magnetoelastic coupling constants for all samples, which is explained in detail in our previous investigation [19]. Including the elastic energy and magnetoelastic coupling we obtain for the Gibbs free energy:

$$\Delta G = \frac{\alpha}{2}M^2 + \frac{\beta}{4}M^4 + \frac{\gamma}{6}M^6 - \mu_0 HM \\ + \frac{1}{2}\sum_{i,j} C_{ij} e_i e_j + \sum_i g_i e_i M \quad (3)$$

where $C_{ij}$ are the elastic constants, $e_i$ is the elastic strain and $g_i$ are the magnetoelastic coupling constants. Considering a hexagonal symmetry, the tensile strains within the basal plane is $e_1 = \Delta a/a$ and along the sixfold axis $e_3 = \Delta c/c$. Minimisation of $\Delta G$ with respect to the strains ($d\Delta G/de_1 = 0$ and $d\Delta G/de_3 = 0$) results in:

$$(C_{11} + C_{12})e_1 + C_{13}e_3 + g_1 M = 0 \quad (4)$$

$$2C_{13}e_1 + C_{33}e_3 + g_3 M = 0 \quad (5)$$

The strain as function of magnetization is shown in Fig. 8(a). The volume change across the transition is tuned to be small (see Fig. 3(n)). Therefore, we can simplify the analysis by defining that $e \equiv e_1 = -\frac{1}{2}e_3$ (see also [19]). By using the derivatives $de_1/dM$ and $de_3/dM$ (Fig. 8(b)) and taking the elastic constants from [24], we can calculate the magnetoelastic coupling constants within the plane $g_1 = -(C_{11}+C_{12}-2C_{13})\, de_1/dM$ and along the sixfold axis $g_3 = -(C_{33} - C_{13})\, de_3/dM$ shown in Fig. 8(c). The elastic constants are $C_{11} = 308.4$, $C_{12} = 120.0$, $C_{13} = 144.0$, $C_{33} = 227.8$, $C_{44} = 119.5$ and $C_{66} = 94.2$ GPa [24].

The strongest magnetoelastic coupling leading to the highest energy gain in the Gibbs free energy is present in the sample without B ($w = 0$), which has the strongest FOMT character. The magnetoelastic coupling constants gradually change and values approach each other when the samples are tuned with an increasing B content towards the SOMT. In contrast to the electron density difference, the sample without B follows the same trend as the other samples containing B for the magneto-elastic coupling. The average of the magnetoelastic coupling constants for all samples in the series are $g_1 = -0.023 \times 10^9$ JkgA$^{-1}$m$^{-5}$ and $g_3 = 0.027 \times 10^9$ JkgA$^{-1}$m$^{-5}$ with an experimental ratio of $\left(\frac{g_3}{g_1}\right)_{exp} = -1.13$. This experimental ratio is in close agreement with the theoretical value for $e_1 = -\frac{1}{2}e_3$ predicted by the Landau model: $\left(\frac{g_3}{g_1}\right)_{th} = -2(C_{33} - C_{13})/(C_{11} + C_{12} - 2C_{13}) = -1.19$.

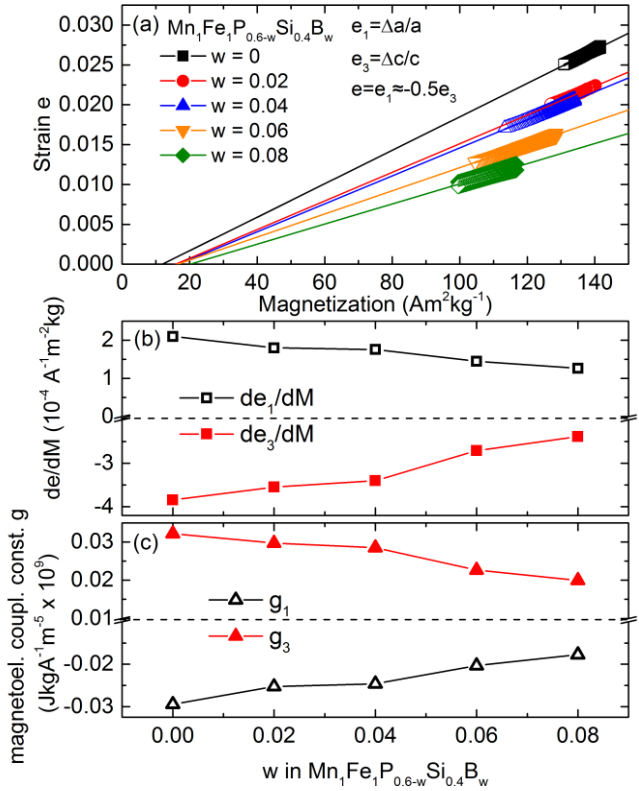

Fig. 8: (a) Strain $e = e_1 = -\frac{1}{2}e_3$ with $e_1 = \frac{\Delta a}{a}$ and $e_3 = \frac{\Delta c}{c}$ in the ferromagnetic phase of $Mn_1Fe_1P_{0.6-w}Si_{0.4}B_w$ ($w = 0 – 0.08$) as function of the magnetization. From (a) linear fits indicated by solid lines we obtain (b) the slope $de_{1,3}/dM$, which is used to calculate (c) the magnetoelastic coupling constants $g_1$ (in $ab$-plane) and $g_3$ (out of plane along $c$-axis). Dashed black lines in (b) and (c) mark breaks in the vertical axis.

For all samples ($w = 0 - 0.08$), we observe similar gradual trends between the change of lattice parameters, the magnetoelastic coupling, the transition temperature $T_C$ and thermal hysteresis (Fig. 7). Our investigation indicates that the magnetoelastic coupling strength is closely linked to the transition temperature and hysteresis in this series. By controlling the B content the jump in lattice parameters across $T_C$ can be tuned to minimize the hysteresis and improve the mechanical properties.

## V. CONCLUSIONS

By systematically increasing the B content, the magnetization, thermal hysteresis, lattice constants and internal coordinates of Mn and Fe consistently evolve. Localized changes in electron density at the Fe-site, indicate bond formation. Electron density changes across $T_C$ are more evenly distributed on the Mn-site. The sample without B shows significantly larger variations in the electron density compared to all samples with B substitution. Furthermore, at high temperatures the changes in electronic structure across $T_C$ increase in this sample, indicating the development of short-range order. An increasing B content leads to increasing changes in the electron density across the transition on the Mn site. On the Fe site, changes are similar in all B samples. We observe a clear trend, that a decreasing jump in the lattice parameters across the magnetic transition enhances $T_C$ and decreases hysteresis.

## ACKNOWLEDGMENTS
The authors thank the beamline staff at the Swiss-Norwegian Beam Lines (BM01) at ESRF and Anton Lefering for their technical support. This work was financially supported by TKI Urban Energy with project number 163GM06.

# Charge redistribution and the Magnetoelastic transition across the first-order magnetic transition in (Mn,Fe)$_2$(P,Si,B)

M. Maschek[1], X. You[1], M. F. J. Boeije[1], D. Chernyshov[2], N. H. van Dijk[1] and E. Brück[1]

[1] Fundamental Aspects of Materials and Energy, Faculty of Applied Sciences, Delft University of Technology, Mekelweg 15, 2629 JB Delft, The Netherlands

[2] BM01, Swiss-Norwegian Beam Lines, European Synchrotron Radiation Facility, 71 avenue des Martyrs, 38000, Grenoble, France

**Supplementary materials**

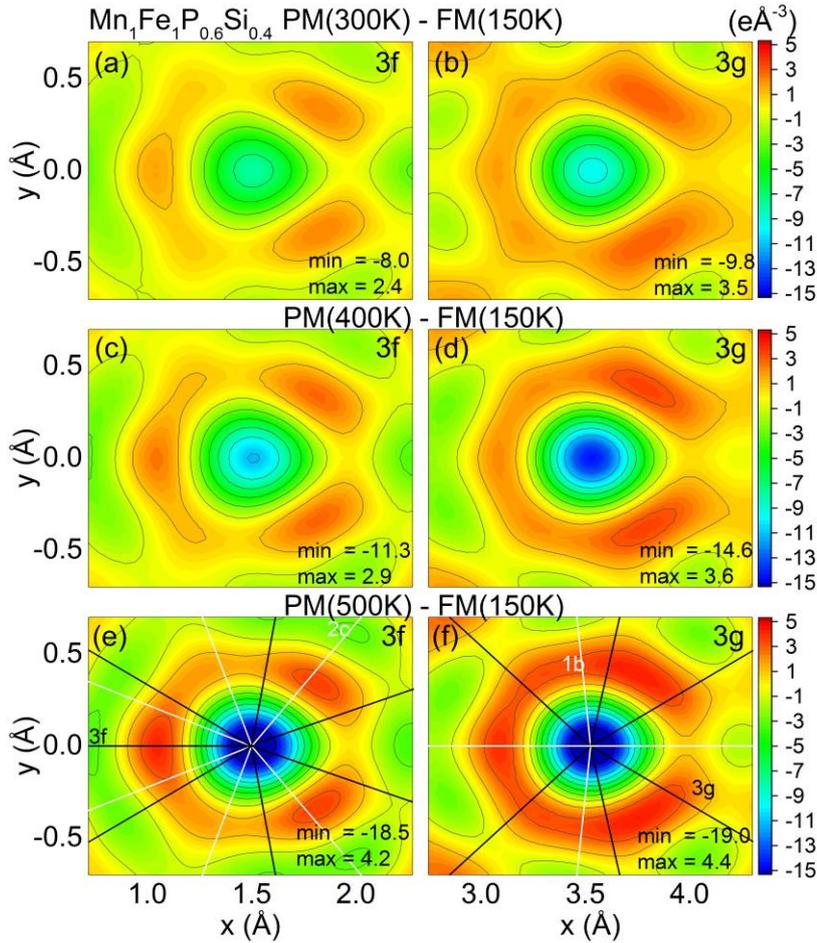

Fig. S1: Electron density difference plots between paramagnetic phase at (a-b) 300 K, (c-d) 400 K, (e-f) 500 K and ferromagnetic phase at 150 K in the local environment of the *3f* and *3g* sites for Mn$_1$Fe$_1$P$_{0.6}$Si$_{0.4}$. Electron densities were integrated from $z = -0.25$ to $0.25$ and $z = 0.25$ to $0.75$ (internal coordinates) for *3f* and *3g* positions, respectively. The shift of internal coordinates with temperature is removed by the corresponding shifts from Fig. 4. in order to emphasize on the effects originating from redistribution of electron densities. The color-coded scale is given in $e$Å$^{-3}$.

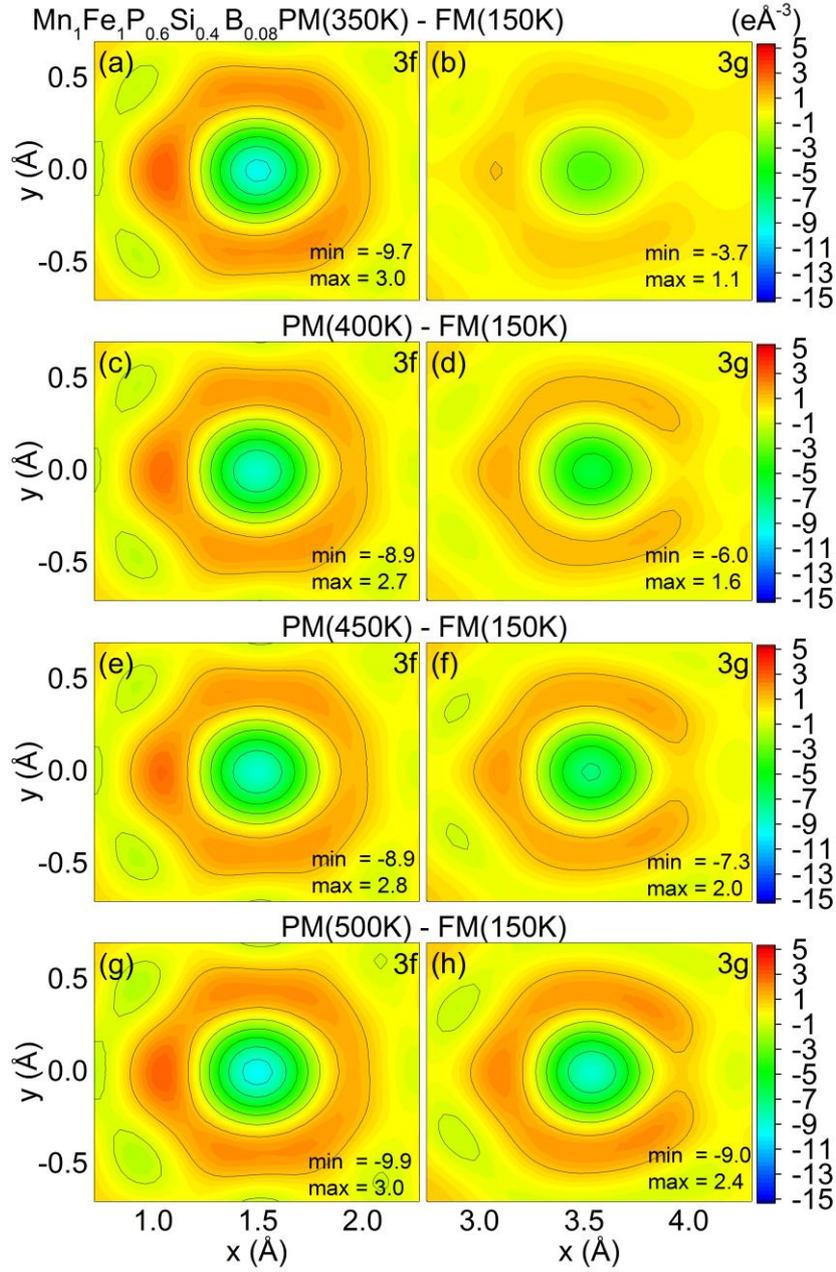

Fig. S2: Electron density difference plots between paramagnetic phase at (a-b) 350 K, (c-d) 400 K, (e-f) 450 K, (g-h) 500 K and ferromagnetic phase at 150 K in the local environment of the *3f* and *3g* sites for $Mn_1Fe_1P_{0.6}Si_{0.4}B_{0.08}$. Electron densities were integrated from $z = -0.25$ to $0.25$ and $z = 0.25$ to $0.75$ (internal coordinates) for *3f* and *3g* positions, respectively. The shift of internal coordinates with temperature is removed by the corresponding shifts from Fig. 4. in order to emphasize on the effects originating from redistribution of electron densities. The color-coded scale is given in $e$Å$^{-3}$.